\documentclass[12pt]{article}
\usepackage{amsmath,amsfonts,amssymb,latexsym}
\newcommand{\newc}{\newcommand}

\newc{\be}{\begin{equation}}
\newc{\ee}{\end{equation}}
\def\bq{\begin{eqnarray}}
\def\eq{\end{eqnarray}}
\def\beq{\begin{eqnarray*}}
\def\eeq{\end{eqnarray*}}
\newc{\ie}{{\it i.e.} }
\newc{\eg}{{\it e.g.} }
\newc{\etc}{{\it etc.} }
\newc{\etal}{{\it et al.}}

\newtheorem{theorem}{Theorem}

\newc{\ra}{\rightarrow}
\newc{\lra}{\leftrightarrow}
\newc{\lsim}{\buildrel{<}\over{\sim}}
\newc{\gsim}{\buildrel{>}\over{\sim}}

\def\simless{\mathbin{\lower 1pt\hbox
   {$\spose{\raise 5pt\hbox{$\char'074$}}\char'430$}}}
\def\simgreat{\mathbin{\lower 1pt\h   {$\spose{\raise 5pt\hbox{$\char'076$}}\char'430$}}}
\def\simgreat{\gapp}
\def\simless{\lapp}
\def\lapp{\mathbin{\raise2pt \hbox{$<$} \hskip-9pt \lower4pt \hbox{$\sim$}}}
\def\gapp{\mathbin{\raise2pt \hbox{$>$} \hskip-9pt \lower4pt \hbox{$\sim$}}}
\newcommand{\rfnce}{\par \noindent \hangindent 15pt{}}
\begin{document}
\setcounter{page}{1}

\title
{\large \bf Future Singularities and Completeness in Cosmology\footnote{Based on a talk presented at the {\it Meeting on "Cosmology and
Gravitational Physics"},  December 2005, Thessaloniki, Greece, to appear in the Proceedings, {\it
Editors:} N.K. Spyrou, N. Stergioulas and C.G. Tsagas.}}

\author {\small \bf \underline {Spiros Cotsakis}$^1$\thanks{skot@aegean.gr}\\ \small $^1$ University of the Aegean,
Karlovassi 83200, Samos, Greece}

\date{}
\maketitle


\begin{abstract}
We review recent work on the existence and nature of cosmological singularities  that can be formed during the evolution of generic as well as specific cosmological spacetimes  in general relativity. We first  discuss necessary and sufficient conditions for the existence of geodesically incomplete spacetimes based on a tensorial analysis of the geodesic equations. We then classify the possible singularities of isotropic globally hyperbolic universes using  the Bel-Robinson slice energy that closely monitors the asymptotic properties of fields near the singularity. This classification includes all known forms of spacetime singularities in isotropic universes and also predicts new types. 

\end{abstract}

\vspace{1cm}

{\bf 1. Introduction}\\
The general issue that we address in this paper is probably the most common one in cosmology: 
\begin{center}Does the universe exist forever? \end{center}
In other words, was there a finite proper time in the past common to whole space before which the universe did not exist? How about in the future? Commonly, if yes, we say the universe has past (resp. future) singularity, otherwise the universe is \emph{complete}. Thus completeness is the negation of singularity and vice versa. In order to avoid endless philosophical debates for even the simplest issues, perhaps the first thing one should do in a subject of this sort  is to have precise mathematical definitions and criteria.

The standard definition of  spacetime completeness in general relativity is to say that spacetime is complete if it is geodesically complete that is every causal (i.e., timelike or null) geodesic defined in a finite subinterval can be extended to the whole real line. Otherwise, spacetime is singular that is geodesically incomplete, there is at least one incomplete geodesic.

The traditional criteria (cf. Hawking and Ellis (1973)) provide sufficient conditions for a spacetime to be singular, that is geodesically incomplete. These conditions basically depend on the effect of spacetime curvature on a \emph{congruence} of causal geodesics, in particular, they predict a blow up singularity  in the expansion of the said congruence a finite parameter value after the congruence has started to converge, that is all geodesics will meet at a point conjugate to what can be called the \emph{surface of last convergence}. All this follows from an analysis of the equation satisfied by the expansion of the bunch of geodesics called the Komar-Landau-Raychaudhouri equation.

A more analytic approach to the singularities of general relativity is to consider directly the equation satisfied not by a family of geodesics as above but by each individual geodesic, the geodesic equation. This is the approach taken in Choquet-Bruhat \& Cotsakis (2002) wherein sufficient analytic conditions were found for geodesics to have infinite proper length. It is important to point out that in this approach the causal techniques, which are used in the traditional singularity theorems, are also used here, especially global hyperbolicity,  however, the tensor-analytic technique is completely different. 

The plan of this paper is as follows. In the next Section, we review the beginnings of the new approach to the singularity problem and state three theorems, the first giving conditions for the equivalence of global hyperbolicity to slice completeness, another giving sufficient conditions for a spacetime to be g-complete (theorrem 3) and another (Theorem 2) offering a converse in the particular case of the so-called trivially sliced spaces. In Section 3, we state a new singularity theorem, that is \emph{necessary} conditions for causal g-incompleteness, of particular relevance to cosmology which allows one to classify  all possible {\textsc FRW} future singularities as being those with a non-integrable Hubble parameter. We also provide examples from the recent literature on sudden singularities and inflationary cosmology to illustrate our results. In Section 4, we introduce the Bel-Robinson energy as a new way towards a refined classification of the possible singularities in the isotropic category. This leads to many different types of singularities, for instance those in tahyonic cosmology. We provide examples which demonstrate the relevance of  many of these new types. Lastly,  we point out further consequences of our work and possible future extensions.

{\bf 2. Geodesics and singularities: Necessary conditions}\\
In this Section, we start with a very general situation. Consider a spacetime of the form $(\mathcal{V},g)$ with
$\mathcal{V}=\mathcal{M} \times \mathcal{I},\;$ $\mathcal{I}$
being an interval in $\mathbb{R}$ (for simplicity we can take it to be the whole real line) and $\mathcal{M}$  a smooth
manifold of dimension $n$, in which the smooth,
$(n+1)$--dimensional, Lorentzian metric $g$ splits as follows:
\begin{equation}
g\equiv -N^{2}(\theta ^{0})^{2}+g_{ij}\;\theta ^{i}\theta
^{j},\quad \theta ^{0}=dt,\quad \theta ^{i}\equiv dx^{i}+\beta
^{i}dt.  \label{2.1}
\end{equation}
Here $N=N(t,x^{i})$ denotes the \emph{lapse function}, $\beta
^{i}(t,x^{j})$ is the \emph{shift function} and the spatial
slices $\mathcal{M}_{t}\,(=\mathcal{M}\times \{t\})$ are spacelike
submanifolds endowed with the time-dependent spatial metric
$g_{t}\equiv g_{ij}dx^{i}dx^{j}$. We call such a spacetime a
\emph{sliced space}, as in Cotsakis (2004). We shall assume that our sliced space $(\mathcal{V},g)$ is \emph{regularly hyperbolic}, that is the lapse function is bounded, the shift is uniformly bounded and the spatial metric is itself uniformly bounded below. Under these conditions we can prove (cf. Choquet-Bruhat \& Cotsakis (2002), Cotsakis (2004)):
\begin{theorem}
Let  $(\mathcal{V},g)$ be a regularly hyperbolic sliced space. Then the
following are equivalent:
\begin{enumerate}
\item $(\mathcal{M}_{0},g_{0} )$ is a g-complete Riemannian manifold
\item The spacetime $(\mathcal{V},g)$ is globally hyperbolic
\end{enumerate}
\end{theorem}
This theorem shows that in a reasonable spacetime the basic causal notion of global hyperbolicity is in fact equivalent to the condition that each slice is a geodesically complete manifold.

Under
what conditions is global hyperbolicity equivalent to geodesic
completeness in the original spacetime $(\mathcal{V},g)$? What is the class of sliced spaces in which such an
equivalence holds? In a sliced space belonging to this class, in
view of the results of the previous Theorem, geodesic completeness
of the spacetime would be guessed simply by 
looking at the completeness of a slice. Let us define a \emph{trivially sliced space} to be a spacetime with constant lapse and shift and time-independent spatial metric. Then we can prove the following result (Cotsakis (2004)).
\begin{theorem}
Let  $(\mathcal{V},g)$ be a trivially sliced space. Then the
following are equivalent:
\begin{enumerate}
\item The spacetime $(\mathcal{V},g)$ is timelike and null geodesically complete
\item $(\mathcal{M}_{0},g_0 )$ is a complete Riemannian manifold
\item The spacetime $(\mathcal{V},g)$ is globally hyperbolic.
\end{enumerate}
\end{theorem}
We see that this theorem gives sufficient and necessary conditions for g-completeness in spacetimes of a particularly simple form, trivially sliced spaces. For more general situations, things can become very complicated and no general results offering sufficient and necessary conditions exist. 

The first result giving \emph{sufficient} conditions for a spacetime to be g-complete was proved in Choquet-Bruhat \& Cotsakis (2002).  The method of proof consisted of a tensor analytic argument using the geodesic equations. The tangent vector $u$ to a geodesic in $(\mathcal{V},g)$  parametrized by arc length,
or by the canonical parameter in the case of a null geodesic, with
components $dx^{\alpha}/ds$ in the natural frame,  satisfies in an
arbitrary frame the differential equations,
\begin{equation}
u^{\alpha }\nabla _{\alpha }u^{\beta }\equiv u^{\alpha }\partial _{\alpha
}u^{\beta }+\omega _{\alpha \gamma }^{\beta }u^{\alpha }u^{\gamma }=0.
\end{equation}
Since $u^{0}\equiv dt/ds$, the zero component of the geodesic equations  can be
written in the form,
\begin{equation}
\frac{d}{dt}\left(\frac{dt}{ds}\right)+\frac{dt}{ds}\left(\omega
_{00}^{0}+2\omega _{0i}^{0}v^{i}+\omega _{ij}^{0}v^{i}v^{j}\right)=0,
\label{3.7}
\end{equation}
where we have set,
\begin{equation}
v^{i}=\frac{dx^{i}}{dt}+\beta ^{i}.
\end{equation}
Setting $dt/ds=y$ and using the standard expressions for the connection coefficients, the geodesic equations yield

\begin{equation}
\log \frac{y(t)}{y(t_{1})}=\int_{t_{1}}^{t}N^{-1}\left(-\partial
_{0}N-2\partial _{i}Nv^{i}+K_{ij}v^{i}v^{j}\right) dt,
\end{equation}
or, using regular hyperbolicity,
\begin{equation}
\log\frac{y(t)}{y(t_{1})}\leq 2\log
N_{m}^{-1}+N_{m}^{-1}\int_{t_{1}}^{t}\left(|\nabla
N|_{g}N_{M}+|K|_{g}N_{M}^{2}\right) dt.
\end{equation}
Since the length (or canonical parameter extension) of the curve $C$ is
\begin{equation}
\int_{t_{1}}^{+\infty }\frac{ds}{dt}\,dt
\end{equation}
and will be infinite if $ds/dt$ is bounded away from zero i.e., if $y\equiv
dt/ds$ is uniformly bounded, we arrive at the following consequence.
\begin{theorem}
If the spacetime $(\mathcal{V},g)$ is globally and regularly hyperbolic and in addition the $g$-norms of the space gradient of the lapse, $|\nabla N|_{g}$,  and of the  extrinsic curvature,$|K|_{g}$, are integrable functions on the interval $[t_{1},+\infty )$, then  spacetime is future timelike and null geodesic
complete.
\end{theorem}
Note that in this theorem, the fact that the length of the geodesics is infinite is proved directly and not using the equation for the expansion of the geodesic congruence as one does to prove the usual singularity theorems. 

{\bf 3. Hubble rate and singularities}\\
The completeness Theorem 2 of the previous Section gives
sufficient conditions for timelike and null geodesic completeness
and therefore implies that the negations of each one of its conditions
are \emph{necessary} conditions for the existence of
singularities (while, the singularity
theorems (cf. Hawking and Ellis (1973)) provide sufficient conditions for this
purpose). Thus this theorem  is precisely what is needed for applications to the  analysis of the nature of specific cosmological models known from other reasons to have spacetime singularities. This is the line of thought followed in Cotsakis and Klaoudatou (2005).

Since for an FRW metric,   $N=1$ and $\beta =0$, we see that regular hyperbolicity is satisfied provided the scale factor $a(t)$ is a bounded from below function of the
proper time $t$ on $\mathcal{I}$. The condition for the integrability of the spatial gradient of the lapse is trivially
satisfied and this is again true in more general
homogeneous cosmologies where the lapse function $N$ depends only
on the time and not on the space variables. In all these cases the norm $|\nabla N|_{g}$ is zero. On the other hand, the condition on the norm of the extrinsic curvature is the
only one which can create a problem. 

For an isotropic universe
 ${|K|_{g}} ^{2}=3({\dot{a}/a})^{2}=3H^{2}$, and
so  FRW universes in which  the scale factor is
bounded below can fail to be complete only when that norm is not integrable.
This can happen in only one way: There is a finite time $t_1$ for
which $H$ fails to be integrable on the time interval
$[t_1,\infty)$. Since this non-integrability of $H$ can be
implemented in different ways, we arrive at the following result
for the types of future singularities that can occur in isotropic
universes (cf. Cotsakis and Klaoudatou (2005).
\begin{theorem}\label{2}
Necessary conditions for the existence of future singularities in
globally hyperbolic, regularly hyperbolic FRW universes are:
\begin{description}
\item[S1] For each finite $t$, H is non-integrable on $[t_1,t]$,
or
\item[S2] H blows up in a finite time, or
\item[S3] H is defined and integrable (that is bounded, finite) for only a finite
proper time interval.
\end{description}
\end{theorem}
When does Condition $S1$ hold? It is well known that a function
$H(\tau)$ is integrable on an interval $[t_1,t]$ if $H(\tau)$ is
defined on $[t_1,t]$, is continuous on $(t_1,t)$ and the limits
$\lim_{\tau\rightarrow t_1^+}H(\tau)$ and $\lim_{\tau\rightarrow
t^-}H(\tau)$ exist. Therefore there are a number of different ways
which can lead to a singularity of the type $S1$ and such
singularities are  in a sense more subtle than the usual ones
predicted by the singularity theorems. 
For instance,   they may
correspond to the so-called \emph{sudden}, or spontaneous,  singularities (see Barrow (2004) for this
terminology) located at the right end (say $t_s$) at which  $H$ is
defined and finite but \emph{the left limit},
$\lim_{\tau\rightarrow t_1^+}H(\tau)$, may fail to exist, thus
making $H$ non-integrable on $[t_1,t_s]$, for \emph{any} finite
$t_s$ (which is of course arbitrary but fixed from the start). 

In  such a model,
we have a solution of the Friedmann equations for a  fluid source
with unconnected density and pressure which, in a local neighborhood of the
singularity located at the time $t=t_s$ ahead, reads
\begin{equation}\label{barrow}
a(t)=1+\left(\frac{t}{a_{s}}\right)^{q}
(a_{s}-1)+\tau^n\Psi(\tau),\quad \tau=t_s-t.
\end{equation}
Here we take $1<n<2,\, 0<q< 1, a(t_s)=a_s$ and $\Psi(\tau)$ is the
so-called \emph{logarithmic psi-series} which is assumed to be
convergent, tending to zero as $\tau\rightarrow 0$. 
The form (\ref{barrow}) exists as a
\emph{smooth} solution only on the interval $(0,t_{s})$. Also
$a_{s}$ and $H_{s}\equiv H(t_s)$ are finite  but
$\dot{a}$ blows up as $t\rightarrow 0$ making $H$ continuous only
on $(0,t_s)$. In addition,  $a(0)$ is finite and we can extend $H$
and define it to be finite also at $0$, $H(0)\equiv H_0$, so that
$H$ is defined on $[0,t_s]$. However, since $\lim_{t\rightarrow
0^+}H(t)=\pm\infty$, we conclude that this model universe
implements exactly Condition $S1$ of the previous Section and thus
$H$ is non-integrable on $[0,t_s]$, $t_s$ arbitrary. 

This then provides an example of a
singularity characterized by the fact that as $t\rightarrow t_{s}$, $\ddot{a}\rightarrow-\infty$ while using the field
equation we see that this is really a divergence in the pressure,
$p\rightarrow\infty$. In particular, we cannot have in this
universe a family of privileged observers each having an infinite
proper time and finite $H$. A further calculation shows that the
product $E_{\alpha\beta}E^{\alpha\beta}$, $E_{\alpha\beta}$ being
the Einstein tensor, is unbounded at $t_s$. Hence we find that
this spacetime is future geodesically incomplete.

As another example consider  a flat FRW  model,
$
ds^{2}=dt^{2}-a^{2}(t)d\bar{x}^{2},
$
and take all quantities along a null geodesic with affine parameter $\lambda$.
Since this is conformally Minkowski we have
$
d\lambda \propto a(t) dt,
$
or,
$
d\lambda=a(t)dt/a(t_{s}),
$
so that $d\lambda/dt=1$ for $t=t_{s}$, where $t_{s}$ is a finite
value of time. Borde {\emph{et al} (2003) then define an averaged-out Hubble rate in this inflating spacetime by the equation 
\begin{equation}
H_{av}=\frac{1}{\lambda(t_{s})-\lambda(t_{i})}
\int_{\lambda(t_{i})}^{\lambda(t_{s})}H(\lambda)d\lambda,
\end{equation}
and so  if $H_{av}>0$, as we would expect in a truly inflationary universe,  we have, following Cotsakis and Klaoudatou (2005),
\begin{equation}\label{7.5}
0<H_{av}=\frac{1}{\lambda(t_{s})-\lambda(t_{i})}
\int_{\lambda(t_{i})}^{\lambda(t_{s})}H(\lambda)d\lambda
=\frac{1}{\lambda(t_{s})-\lambda(t_{i})}
\int_{a(t_{i})}^{a(t_{s})}\frac{da}{a(t_{s})}\leq\frac{1}
{\lambda(t_{s})-\lambda(t_{i})}.
\end{equation}
This shows that the affine parameter must take values only in a finite interval (otherwise we would get a vanishing $H$)
which implies geodesic incompleteness. A similar proof is obtained for the case of a timelike
geodesic. Notice that condition (\ref{7.5}) holds if and only if
$H$ is integrable  on only a finite interval of time, thus
signalling incompleteness according to Theorem 4. 

{\bf 4. Bel-Robinson energy and the nature of singularities}\\
Instead of starting from the geodesic equation, we take in this Section another path, a tensorial in spirit  approach, and consider the Bel-Robinson energy (cf. Cotsakis and Klaoudatou (2006)). In a sliced space this can be defined as follows.  We start with the $2$-covariant spatial electric
and magnetic tensors, defined by (cf. Choquet-Bruhat \& York (2002)
\beq
E_{ij}&=&R^{0}_{i0j},\\
D_{ij}&=&\frac{1}{4}\eta_{ihk}\eta_{jlm}R^{hklm},\\
H_{ij}&=&\frac{1}{2}N^{-1}\eta_{ihk}R^{hk}_{0j},\\
B_{ji}&=&\frac{1}{2}N^{-1}\eta_{ihk}R^{hk}_{0j},
\eeq
where
$\eta_{ijk}$ is the volume element of the space metric $ g$.
The \emph{Bel-Robinson energy} is then defined as the slice integral 
\be
\mathcal{B}(t)=\frac{1}{2}\int_{\mathcal{M}_t}\left(|E|^{2}+|D|^{2}+
|B|^{2}+|H|^{2}\right)d\mu_{\bar{g}_t}, \ee where by
$|X|^{2}=g^{ij}g^{kl}X_{ik}X_{jl}$ we denote the spatial norm of
the $2$-covariant tensor $X$. The Bel-Robinson energy is a kind of energy of the gravitational
field \emph{projected} in a sense to a slice in spacetime and as such it can be very useful in evolution studies of the Einstein equations. It was in fact 
used in Choquet-Bruhat \& Moncrief (2002) to prove global existence results
for cosmological spacetimes.

For an $\textsc{FRW}$ universe filled with various forms of matter
with metric given by
$ds^2=-dt^2+a^2(t )d\sigma ^2,$
where
$d\sigma ^2$ denotes the usual time-independent metric on the
3-slices of constant curvature $k$, the norms
of the magnetic parts, $|H|, |B|$, are identically zero while $|E|$ and
$|D|$, the norms of the electric parts, reduce to 
\be
|E|^{2}=3\left(\ddot{a}/{a}\right)^{2} \quad\textrm{and}\quad
|D|^{2}=3\left(\left(\dot{a}/{a}\right)^{2}+k/{a^{2}}\right)^{2}.
\ee
Therefore the Bel-Robinson energy becomes
\be \mathcal{B}(t)=\frac{C}{2}\left(|E|^{2}+|D|^{2}\right),
\ee
where $C$ is the constant volume of (or \emph{in} in the case of a
non-closed space) the 3-dimensional slice at time $t$.

The approach  to the singularity problem introduced in Cotsakis and Klaoudatou (2006) classifies the possible types of singularities that are
formed in an $\textsc{FRW}$ geometry during its cosmic evolution using the different possible
combinations of the three main functions in the problem, namely,
the scale factor $a$, the Hubble expansion rate $H$ and  the Bel
Robinson energy $\mathcal{B}.$ If we
suppose that the model has a finite time singularity at $t=t_s$, then
the possible behaviours of the functions in the triplet $\left(H,a,(|E|,|D|)\right)$
in accordance with Theorem 4 are as follows:
\begin{description}
\item [$S_{1}$] $H$ non-integrable on $[t_{1},t]$ for every $t>t_{1}$

\item [$S_{2}$] $H\rightarrow\infty$  at $t_{s}>t_{1}$

\item [$S_{3}$] $H$ otherwise pathological
\end{description}

\begin{description}
\item [$N_{1}$] $a\rightarrow 0$

\item [$N_{2}$] $a\rightarrow a_{s}\neq 0$

\item [$N_{3}$] $a\rightarrow \infty$
\end{description}

\begin{description}
\item [$B_{1}$] $|E|\rightarrow\infty,\, |D|\rightarrow \infty$

\item [$B_{2}$] $|E|<\infty,\, |D|\rightarrow \infty $

\item [$B_{3}$] $|E|\rightarrow\infty,\, |D|< \infty $

\item [$B_{4}$] $|E|<\infty,\, |D|< \infty $.
\end{description}
The nature of a prescribed  singularity is thus described completely by
specifying the components in a triplet of the form $(S_{i},N_{j},B_{l}),$
with the indices $i,j,l$ taking their respective values as above.
Except some impossible cases (cf. Cotsakis and Klaoudatou (2006)), all other types of finite time singularities can in principle be
formed during the evolution of \textsc{FRW}, matter-filled models,
in general relativity or other metric theories of gravity.
It is interesting to note that all the standard dust or
radiation-filled big bang singularities fall under the
\emph{strongest} singularity type, namely, the type
$(S_{1},N_{1},B_{1})$. For example, in a flat universe filled with
dust, at $t=0$ we have \bq a(t)&\propto& t^{2/3}\rightarrow 0,
\quad (N_{1}),\\ H&\propto& t^{-1}\rightarrow\infty, \quad
(S_{1}),\\ |E|^{2}&=&3/4H^{4}\rightarrow\infty,\quad
|D|^{2}=3H^{4}\rightarrow\infty, \quad (B_{1}). \eq Note that this
scheme is organized in such  a way that the character of the
singularities (i.e., the behaviour of the defining functions)
becomes milder as the indices of $S$, $N$ and $B$ increase (this applies to both collapse and rip-type singularities). Milder
singularities in isotropic universes are thus expected to occur as
one proceeds down the singularity list.

Further, we note that we can obtain a more detailed picture of the possible singularity formations by studying the relative asymptotic behaviours of the
three functions that define the type of singularity. Using a
standard notation that expresses the behaviour of two functions
around the singularity at $t_{\ast}$, we  introduce the notion of \emph{
relative strength} in the singularity  classification. Let $f,g$
be two functions. We say that
\begin{enumerate}
\item $f(t)$ is \textit{much smaller } than $g(t)$, $f(t)<<g(t)$, if and only if \\
$\lim_{t\rightarrow t_{\ast}} f(t)/g(t)=0$

\item $f(t)$ is \textit{similar} to $g(t)$, $f(t) \sim  g(t)$, if and only if
$\lim_{t\rightarrow t_{\ast}} f(t)/g(t)<\infty$

\item $f(t)$ is \textit{asymptotic} to $g(t)$, $f(t)  \leftrightarrow g(t)$,
if and only if $\lim_{t\rightarrow t_{\ast}} f(t)/g(t)=1$.
\end{enumerate}
Using these relations
we find that for example, the standard radiation filled isotropic universes (with $k=0,\pm 1$) have the asymptotic behaviours described by
$$a<<H<<(|E|\leftrightarrow |D|),$$ whereas the rest of the standard big
bang singularities have $$a<<H<<(|E| \sim  |D|).$$ This shows the precise nature of these two specific types of  singularities. Many more examples can be found in Cotsakis and Klaoudatou (2006) (see also Cotsakis and Klaoudatou (2006b)).

The Bel-Robinson energy can also be used to test g-completeness. For instance, consider the case of \emph{tachyonic} cosmologies as in Cotsakis and Klaoudatou (2005b). Tachyons, phantoms, Chaplygin gases etc, represent
unobserved and unknown, tensile, negative energy and/or pressure
density substances, violating some or all of the usual energy
conditions. The chief purpose of such exotic types of matter is to cause cosmic acceleration and drive the late phases of the evolution of the universe. However, they are also bound to have a nontrivial effect on the singularity structure of such universes.

A question of interest to us is under what conditions are cosmological models sourced by such fields g-complete?
Following Cotsakis and Klaoudatou (2005b), we consider again the Friedman model filled with a Chaplygin gas with equation of state given by
$
p=-{\rho}^{-\alpha}[C+(\rho^{1+\alpha}-C)^{\alpha/(1+\alpha)}],
$
where $C=A/(1+w)-1$ and subject to the condition
$1+\alpha=1/(1+w)$. The scale factor is then given by the form
$
a(t)=\left(C_{1}e^{-C_3\tau}
+C_{2}e^{C_3\tau}\right)^{2/3}, \tau=t-t_0,
$
where $C_{1}$, $C_{2}$ and $C_{3}$ are constants. Therefore we find that in the asymptotic
limits  $\tau\rightarrow 0$ and $\tau\rightarrow\infty$,
$H$ tends to suitable constants, that is it
remains finite on $[t_{0},\infty)$ and so by our theorems  the model is geodesically complete. 
We can also arrive at the same conclusion using the Bel-Robinson energy of this universe which in this case can be shown that it tends to a constant value thereby signalling completeness. 

However, this is by no means the only possible dynamical behaviour in phantom/tachyonic cosmologies. An example of g-incomplete universe is the so-called graduated inflationary models first considered in  Barrow (1990). A more general class of models can be described by a flat space with a fluid having equation of state $p+\rho=\gamma\rho^{\lambda}$, $\gamma>0$ and $\lambda<1$. This family was further analyzed in Cotsakis and Klaoudatou (2006) using the Bel-Robinson energy. It was found that the exact behaviour  of the model (which  as we showed exhibits an $S2$ type singularity) described originally by Barrow (1990) has a more general significance and the Bel-Robinson energy diverges making this class of models g-incomplete. 

{\bf 5. Discussion}\\
In this paper we have reviewed the recent introduction and application of two new methods to tackle the singularity problem in cosmology. The first method is based on a direct tensorial treatment of the geodesic equations that enables one to prove sufficient conditions for a spacetime to be g-complete, that is conditions under which geodesics have infinite length. This in conjuction with causal techniques allow one to show the equivalence of the deterministic concept of global hyperbolicity to the analytic concept of slice as well as general spacetime completeness. Also in the case of simple cosmological spacetimes one can prove necessary conditions for g-completeness, offering a converse to the completeness theorem. In this sense, a singular  spacetime can then be tested directly as to how it becomes singular near a finite time singularity  by checking which functions blow up and in what way at the finite time singularity. 

This is in sharp contrast to the traditional method of showing geodesic incompleteness in general relativity where one considers the behaviour of a geodesic congruence and indirectly shows that the geodesics in the conguence must under certain geometric conditions have finite length.
The conditions reviewed in this paper allow, in the homogeneous and isotropic case, a complete analysis to be made for the finite time singularities by looking at the behaviour of the Hubble rate (extrinsic curvature). This leads to a classification of the relevant singularities in such models. In fact this approach results in an interesting demarcation and comparison of the singularity structures in different evolution cosmological  models.  

A further advance and refinement in this approach is possible with the consideration of the Bel-Robinson energy of the spacetime in addition to the Hubble rate and the scale factor in such universes. In this second analytic technique the Bel-Robinson energy monitors the asymptotic behaviour of the matter fields and is very sensitive even with slight changes in the matter content. This whole approach has led to a prediction of many different types of finite time singularities in cosmological models and their nature has become more amenable to an analytic treatment.

There are many deep, interesting  and feasible open problems that remain in this field and a future synthesis of causal, tensor, dynamical as well as asymptotic methods will prove very useful in progressing towards the goal of discovering all different possibilities. 

{\bf Acknowledgements}\\
I am very grateful to Professor Spyrou and his group for the excellent organization and  stimulating atmosphere of the Thessaloniki Meeting. This work was supported by the joint Greek Ministry of Education
and European Union research grants `Pythagoras' No. 1351 and `Heracleitus' No. 1337.

\vspace{0.1cm}
{\bf {References}}\\
\rfnce
 Barrow, J.~D., \emph{Sudden future singularities}, Class. Quant.
Grav. 21 (2004) L79-L82 [arXiv: gr-qc/0403084].
\rfnce
Barrow, J. D., \emph{Graduated inflationary universes}, Phys. Lett. B235 (1990) 40-43.

\rfnce
Borde, A., \& \emph{et al}, \emph{Inflationary spacetimes are not past-complete}, Phys. Rev. Lett. {\bf 90} (2003) 151301, [arXiv:gr-qc/0110012\.
\rfnce
Choquet-Bruhat, Y. and  S. Cotsakis, \emph{Global hyperbolicity and
completeness}, J. Geom. Phys. 43 (2002) 345-350 (arXiv:
gr-qc/0201057); see also, Y. Choquet-Bruhat, S. Cotsakis, \emph{Completeness
theorems in general relativity}, in \emph{Recent Developments in
Gravity}, Proceedings of the 10th Hellenic Relativity Conference,
K.~D. Kokkotas and N. Stergioulas eds., (World Scientific 2003),
pp. 145-149.

\rfnce
Choquet-Bruhat, Y. and V. Moncrief, \emph{Non-linear Stability of
an expanding universe with $S^1$ Isometry Group}
(arXiv:gr-qc/0302021).

\rfnce
Choquet-Bruhat, Y. and J. W. York, \emph{Constraints and evolution in cosmology}, In: \emph{Cosmological Crossroads}, S. Cotsakis and E. Papantonopoulos (eds.), LNP592, (Springer, 2002), pp. 29-58.
 
\rfnce Cotsakis, S., \emph{Global hyperbolicity and completeness}, Gen Rel. Grav. 36 (2004) 1183-1188.

\rfnce
Cotsakis, S.,  and I. Klaoudatou, \emph{Future singularities of isotropic cosmologies}, J. Geom. Phys. 55 (2005) 306-315.

\rfnce
Cotsakis S.,  and I. Klaoudatou, (2005b) \emph{Modern approaches to cosmological singularities}, In the Proceedings of the 11th Conference on Recent Developments in Gravity, S. Cotsakis and J. Miritzis (eds.), J. Phys. Conf. Series 8 (2005) 150-154.
\rfnce
Cotsakis, S.,  and I. Klaoudatou, \emph{Cosmological singularities and Bel-Robinson energy}, Samos Preprint RG-MPC/060405-1, [arXiv:gr-qc/0604029].

\rfnce
Cotsakis, S.,  and I. Klaoudatou (2006b), \emph{Singular isotropic cosmologies and Bel-Robinson energy}, to appear in the Proceedings of the A. Einstein Century International Conference, Paris, France, [arXiv:gr-qc/0603130].

\rfnce
 Hawking, S. W. and G.~F.~R. Ellis, \emph{The large-scale structure of
space-time}, Cambridge University Press, 1973.

\end{document}